\documentstyle[aps,pra]{revtex}
\draft
\begin{document}
\title{One Spin-Polaron Problem in the Two-Dimensional Kondo-Lattice}
\author{L.A. Maksimov}
\address{{\normalsize Russian Research Center, Kurchatov Institute Kurchatov sq.46,}\\
Moscow, 123182, Russia.\\
A.F.Barabanov\\
{\normalsize Institute for High Pressure Physics, Troitsk, Moscow region,}\\
142092, Russia,\\
R.O. Kuzian\\
{\normalsize Institute for Materials Science, Krjijanovskogo 3, Kiev,}\\
252180, Ukraine}
\date{}
\maketitle

\begin{abstract}
Within the frameworks of spin-polaron concept and the spherically symmetric
state for the antiferromagnetic spin background, the one-particle motion is
studied for two-dimensional Kondo-lattice. The elemetary excitations are
represented as a Bloch superposition of four one-site electron states: two
local states- a bare electron state and a local spin-polaron of small
radius, and two states of delocalized polarons which correspond to the
coupling of local states to the antiferromagnetic spin wave with momentum $%
Q=(\pi ,\pi)$, so called Q-polarons. As a remarkable result we show that the
lowest band of elementary excitations is essentially determined by Q-polaron
states in strongly coupled regime. The account of Q-polarons shifts the band
bottom from $(\pi ,\pi )$ to $(0,0)$. The spectral weight of a bare particle
in the lowest band states can greatly differ from 1. This may lead to a
large Fermi surface for relatively small particle concentration.
\end{abstract}


In order to understand the nature of high temperature superconductors it is
important to describe properly the motion of a hole in the $CuO_2$ plane 
\cite{dagotto,brenig}. This motion takes place on the antiferromagnetic spin
background of copper spins and it must be treated as a correlated motion of
a hole coupled to spin excitations (a spin polaron). Usually the spin
polaron is studied within the frameworks of the t-J model \cite{dagotto} and
the three-band Hubbard model \cite{emery,zh,eme,bar}. These models seem to
be rather complex because the effective spin-fermion coupling appears there
only through the fermion hopping. Moreover, in the case of t-J model it is
difficult to take into account the local constraint which represents the
strong on-site fermion repulsion. We shall study the formation of the spin
polaron within the frameworks of the more simple Kondo-lattice model. Note,
that a generalized variant of the model leads to description of low-lying
states in the three-band Hubbard model \cite{perlov}.

Treating this model Schrieffer \cite{schrieffer} replaced the spin subsystem
by a classic field with a doubled lattice period and described the
elementary excitations as the superposition of electronic states with the
momenta ${\bf {p}}$ and ${\bf {p+Q}}$, where ${\bf {Q}=(\pi,\pi)}$ is the
antiferromagnetic vector. Such a description is deficient for the
two-dimensional $S=\frac{1}{2}$ antiferromagnet because of strong spin
fluctuations. Moreover, it becomes impossible if the spin subsystem is found
in the homogeneous spherically symmetric state with long range order but
with zero value of the average site spin $<S^{\alpha}>=0$ (the homogeneous
Neel state) \cite{arovas,Shimah,Starykh}. The homogeneous Neel state seems
more adequate, then the usual two-sublattice Neel state, for treating the
ground state of the doped $CuO_2$ plane.

The distinctive feature of the present investigation consists in considering
of the one-electron motion on the background of the homogeneous Neel state
of the Kondo lattice. This motion is described by a spin polaron with the
spectrum periodicity relative to the full Brillouin zone. Note that the
conventional two-sublattice spin approach leads to periodicity relative to
the magnetic Brillouin zone \cite{perlov,schrieffer}.

Another distinctive feature of our investigation consists in treating the
spin-polaron as a complex quasiparticle accounted for a coherent
superposition of a bare electron, of a local polaron and of two types of
antiferromagnetic delocalized polarons. A local polaron can be attributed to
the local singlet (an analogous of Zang-Rice polaron in the three-band
Hubbard model) and such a polaron represents the lowest band in the limit of
the strong Kondo interaction J. As to the antiferromagnetic delocalized
polaron it is a bound state of an electron (or of a local polaron) with a
spin wave with the momentum $Q$. As we know such bound states were not
investigated previously. If the spin subsystem is found in the state with
the antiferromagnetic long range order then the amplitude ${S}_Q$ of the
spin wave with $q=Q$ (the Q-wave) has the macroscopic large value and has
the properties analogous to the amplitude of a Bose particle with the zero
momentum in the superfluid Bose-gas. As a result for many problems this
amplitude can be treated as a $c-$number. Then the coupling of the $Q$-wave
to local electron states does not represent new states but leads, as
mentioned above, to mixing of the states with the momenta $p$ and $p+Q$. But
this treatment fails in the case of the homogeneous Neel state. In this
background the average value $<{S}_Q>=0$ and only $<{S}_Q{S}_Q>$ can be
treated as a macroscopic value. Then the coupling of a particle local state
to ${S}_Q$ corresponds to a new delocalized state. It will be shown that it
is important to take into account such a quantum nature of the spin Q-wave
because the transitions between the local polaron states and the delocalized
polaron states essentially determine the spin polaron bands.

The Kondo-lattice Hamiltonian has the form 
\begin{eqnarray} H^{tot}=H_0 + H_1 +H_2, \label{a} \\
H_0=\sum_{rg} t_{g} a_{r+g}^{+}a_{r} =\sum_p
\epsilon_{p}a_{p}^{+}a_{p}, \label{b} \\ H_1=J\sum_{r}
a_{r}^{+}\tilde S_{r}a_{r}, \label{c}\\
H_2=\frac{1}{2}I\sum_{rg}S_{r+g}^{\alpha}
S_{r}^{\alpha}.\label{d} \end{eqnarray}

Here the sums run over the sites $r$ of a square lattice  and over the
nearest neighbors with the lattice spacing $\mid g\mid =1$. For short we
miss the spin index for creation $a_{r\sigma }^{+}$ and annihilation $%
a_{r\sigma }$ operators of the Fermi particles (we shall name them as
electrons) and in the Hamiltonian of Kondo-interaction $H_1$ we take the
notation $\tilde S_r=S_r^\alpha \sigma ^\alpha $.

Let us represent the first two equations of the infinite chain of equations
for the retarded Green's functions, which describe the motion of one
particle in the antiferromagnetic spin background. In the present paper we
restrict ourself mainly to the analysis of the spectrum of the
quasiparticles. That is why for simplicity of notations we shall simply
write operators $a$ instead of the Green's functions $<<a\mid b^{+}>>$ and
shall miss the nonhomogeneous terms $<[a,b^{+}]_{+}>$. The equations have
the form

\begin{eqnarray} (\omega - \epsilon_{p})a_{p} = Jb_p, \quad
b_p = N^{-\frac{1}{2}}\sum_{r}b_r e^{-ipr}
, \quad b_r =  \tilde S_{r}a_{r}
, \label {dot a} \end{eqnarray}
\begin{eqnarray} \lefteqn{ \omega \tilde
S_{r+R}a_{r} =[\tilde S_{r+R}a_{r},H_0 + H_1 + H_2]  =
}\nonumber \\ && = \sum_{g}t_{g} \tilde
S_{r+R}a_{r+g} +J \tilde S_{r+R}\tilde
S_{r}a_{r} + Ii e_{\alpha \beta
\gamma} \sum_{g}\sigma_{\alpha}
S_{r+R+g}^{\beta} S_{r+R}^{\gamma}
a_{r}. \label{dot Sa} \end{eqnarray}

In Ref.\cite{schrieffer} the mean-field approach was taken for the Neel
lattice and the spins were treated as the classical vectors:

\begin{eqnarray} S_{r}^{\alpha} = \delta_{\alpha z}
S_{0} e^{iQr}, \quad S_0 =const, \quad 
Q=(\pm \pi,\pm \pi).          \label{neel}
\end{eqnarray}

In this approximation the Kondo-interaction Hamiltonian (\ref{c}) takes the
form of the potential energy with the doublet period and the equation (\ref
{dot a}) has a closed form:

\begin{eqnarray}
(\omega -\epsilon_{p})a_{p} =  \sigma J
S_{0}a_{p-Q}. \label{dot sr} \end{eqnarray}

Here $p$ is in the Brillouin zone with the periods $(2\pi ,0), (0,2\pi )$,
but the spectrum has the periodicity of the magnetic Brillouin zone:

\begin{eqnarray} E_{ap}= \pm \sqrt{\epsilon_{p}^2 +
\Delta^2}, \qquad \Delta = S_{0} J. \label{E}
\end{eqnarray}

In (\ref{E}) and below we suppose the spectrum model with the "nesting":

\begin{equation}
\epsilon_{p} = - \epsilon_{p+Q} = -2t(cosp_x + cosp_y ) 
\end{equation}
.

In the case of the $S=\frac 12$ spin system the quantum fluctuations are
important and the equation (\ref{dot Sa}) must be used in order to find the
last term in Eq.(\ref{dot a}). In one's turn the equation (\ref{dot Sa})
hasn't a closed form. In order to close the chain of equations we use the
standard Mori-Zwanzig projection technique \cite{mori}. In our case this
means that we must approximate the last two terms in the right-hand side of (%
\ref{dot
Sa}) by their projections on the restricted space of bases
operators. The choice of the restricted set of basis operators must be
dictated by the physics of the problem under discussion.

For a paramagnetic spin subsystem state the simplest set of basis operators
may be taken as two operators that appear in the first equation (\ref{dot a}%
), that are the operator of a ''bare'' electron $a_r$ and the annihilation
operator of an one-site spin polaron $b_r$.The local polaron excitations in
the Kondo lattice were previously discussed in Ref.\cite{perlov}

Now let the spin subsystem to be found in the homogeneous Neel state with
the tensor parameter of the long range order:

\begin{equation}
\label{ctt}<S_{r}^{\alpha}>=0, 
\end{equation}
\begin{equation}
\label{ctw}C_{R}^{\alpha \beta} = <S_{r}^{\alpha}S_{r+R}^{\beta}>= M^{\alpha
\beta} e^{iQR},\qquad \mid {R} \mid >>1. 
\end{equation}

Then it will be consecutively to introduce the operators which take into
account the correlation of the electron and the local polaron with the
antiferromagnetic $Q$-wave

\begin{equation}
\label{cr}c_{r} = \tilde Q_{r}a_{r},\quad c_{p} = \tilde Q_{0}a_{p+Q}, 
\end{equation}
\begin{equation}
\label{qcr}\tilde Q_r = N^{-1}\sum_{R}e^{iQR}\tilde S_{r+R}=e^{iQR}\tilde Q%
_{0}\: . 
\end{equation}
\begin{equation}
\label{dr}d_{r} = \tilde Q_r \tilde S_{r} a_{r},\quad d_{p} = \tilde Q_{0}
b_{p+Q} . 
\end{equation}

Note, that the operators $c_r$ and $d_r$ allow for distant correlations
between the spin and the electron $(\mid R\mid >>1)$. So, in a sense our
approach is alternative to the widely used t-J model investigations based on
the decoupling of an on-site Fermi operator into a spinless fermion and an
antiferromagnetic magnon operator \cite{varm,KLR,horsch,liu}.

It is substantial that the adopted set of the basis operators is closed
relative to the Kondo-interaction Hamiltonian: 
\begin{eqnarray}
\big[a_{r}, H_{1}\big] & =&  J b_{r}\:, \nonumber \\
\big[b_{r}, H_{1}\big] & =&  J(\frac{3}{4}a_{r} - b_{r})\:,\nonumber \\
\big[c_{r}, H_{1}\big] & =&  Jd_{r} \:, \label{QQ} \\
\big[d_{r}, H_{1}\big] & =&  J(\frac{3}{4}c_{r} - d_{r} )\:.
\nonumber
\end{eqnarray}

Taking into account Eq.(\ref{dot Sa}) this gives,in particular, 
\begin{equation}
\label{dot c}(\omega - \epsilon_{p+Q})c_{p} = Jd_p. 
\end{equation}

In order to get a closed form for the Green's functions equations we shall
project the corresponding commutators on the following orthonormal set of
basis operators:

\begin{eqnarray}
B_{1r} & = & a_{r} \:, \nonumber \\
B_{2r} & = & \frac{1}{\sqrt{f_{2}}}b'_{r},  \quad
b'_{r} = b_{r} -c_{r} \: , \nonumber \\
f_{2}  & = & <[b'_{r},b'_{r}{}^{+}]_{+}>=
\frac{3}{4}-M, \quad M=M^{\alpha\alpha} \:,\nonumber  \\
B_{3r} & = & \frac{1}{\sqrt{f_{3}}} c_{r} \: ,\label{BB}  \\
f_{3}  & = & <[c_{r} ,c^{+}_{r} ]_{+}> = M \:, \nonumber \\
B_{4r} & = & \frac{1}{\sqrt{f_{4}}}d'_{r},  \quad
d'_{r} = d_{r} -Ma_{r} +
\frac {M} {f_{2}}(b_{r} - c_{r}) \: , \nonumber \\
f_{4}  & = & <[d'_{r},d'^{+}_{r}]_{+} >
= M(f_{2}-\frac{M}{f_{2}}). \nonumber 
\end{eqnarray}

The coefficients of the projection relations
\begin{equation}
\label{pp}\big[B_i,H\big]=\sum_ja_{ij}B_j
\end{equation}
are determined as

\begin{equation}
\label{aij}a_{ij} = <\big[[B_i,H],B_{jr}^{+}\big]_{+}>. 
\end{equation}

Taking into account the equations (\ref{dot 
a}), (\ref{dot c}) and (\ref
{aij}), we obtain in momentum representation a simple system of four
equations 
\begin{equation}
\label{pup}\omega B_{i,p} = \sum_{j}a_{ij}B_{j,p} 
\end{equation}

The matrix elements of the spectral matrix $a_{ij}$ are equal 
\begin{eqnarray}
a(1,1)=\epsilon_p , \quad   a(1,2)=J\sqrt{f_2},
\quad     a(1,3)=J\sqrt{M}, \quad  a(1,4)=0, \nonumber    \\
a(2,2) = \frac{1}{f_2}\big[ (C+M)\epsilon_p -
 J(f_2 -M)-4IC \big], \nonumber \\
a(2,3) = -J\sqrt{\frac{M}{f_2}}, \label{M} \\
a(2,4) = \big[ \frac{M}{f_2}(C+M)\epsilon_p - Jf_4 - 2IM
(1 + \frac{2C}{f_2}) \big]\frac{1}{\sqrt{f_2 f_4}}, \nonumber \\
a(3,3) = - \epsilon_p,   \quad
a(3,4) = J\sqrt{\frac{f_4}{M}}, \nonumber \\
a(4,4) = \frac{M}{f_4}
\big[ (C + M)\frac{M}{f_2^2} - (C+\frac{M}{3}) \big]
\epsilon_p
- \frac{3J}{4f_2} +   
\frac{IM}{f_4}\big[\frac{8M}{3} - 4C(1 + \frac{M}{f_2^2})
- \frac{4M}{f_2}\big].
\nonumber \end{eqnarray}

Here $C=C_{R=1}^{\alpha \alpha }<0$. To obtain these expressions we used the
following approximation of Takahashi \cite{Tak} for four different-site spin
correlation functions

\begin{equation}
<S_{r_1}^i S_{r_2}^j S_{r_3}^k S_{r_4}^l > =
C_{r_{1}-r_{2}}^{ij}C_{r_{3}-r_{4}}^{kl}+
C_{r_{1}-r_{3}}^{ik}C_{r_{2}-r_{4}}^{jl}+
C_{r_{1}-r_{4}}^{il}C_{r_{2}-r_{3}}^{jk}, r_1\ne r_2\ne r_3\ne r_4. 
\nonumber 
\end{equation}

Hence, in the adopted approximation the electron motion in the
antiferromagnetic background is described by the quasiparticle that is a
coherent superposition of four Fermi fields - the field of the bare electron 
$a_p $, the field of the delocalized polaron $c_p $ and two fields of the
localized polaron $b_p$ and $d_p$, which are hybridized mainly due to the
Kondo-interaction. The system of equations (\ref{pup}) can be treated as the
system of the Schroedinger equations where the matrix elements $a_{ij}$
reproduce the amplitude of the transitions from the state $j$ to the state $%
i $. The presence of the nondiagonal matrix elements underlines the quantum
nature of the spin $S$ and the $Q$-waves. The eigenfunctions and the
eigenvalues of Eqs. (\ref{pup}) describe the elementary excitations which
generate four bands. The form of the bands depends on the state of the
magnetic subsystem (the quantities $M$ and $C$) and the relations between
the energetic parameters $t, J, I $. Below we take the following typical
values for the quantities $M$ and $C$: $M=0.1$ and $C=-0.335$ \cite
{man,Shimah,Starykh}.

In the adopted model all the spectra depend on momentum only through $%
\epsilon _p$. Let us mention that calculating the matrix elements (\ref{aij}%
) we took the approximation of low density of Fermi particles, i.e., $%
<a_p^{+}a_p>\to 0$.

It is essential that the resulting four bands are only approximately
symmetric relative to the magnetic Brillouin zone boundary. In the momentum
space the distance between two quasiparticle states with the equal energy
can differ from the antiferromagnetic vector $Q$. This means that dictated
by these bands the ''shadow band'' effect can be other than the usual one
displaced by $Q$ vector in the two-sublattice Neel antiferromagnet \cite
{Kampf}.

Four waves of the quasiparticle spectrum $E^{(i)}(p),i=1-4$, given by the
eigenvalues of the system of equations (\ref{pup}), have a clear physical
meaning in the cases when one of energetic parameters $t,J,I$ is much
greater then the others. The bands for these cases are represented in Figs.
1-3(a). In Figs. 1-3(b) we also represent the values of the electron Green's
function residues $Z_p^{(i)}$ which correspond to the poles $E^{(i)}(p)$ 
\begin{equation}
\label{Zp}<<a_{p\sigma }\mid a_{p\sigma }^{+}>>=<<B_{1,p}\mid
B_{1,p}^{+}>>=\sum_{i=1}^4\frac{Z_p^{(i)}}{\omega -E^{(i)}(p)}.
\end{equation}
In the low density limit the residues $Z_p^{(i)}$ are determined by the
solution of the equations of motion for Green's functions $<<B_{i,p}\mid
B_{1,p}^{+}>>$ when the nonhomogeneous terms are equal to $%
W_{i,1}=<[B_{i,p},B_{1,p}^{+}]_{+}>=\delta _{i,1}$. The $Z_p^{(i)}$ values
represent one-particle spectral function $A(p,\omega )=\sum_iZ_p^{(i)}\delta
[\omega -E^{(i)}(p)]$. Each of these values $Z_p^{(i)}$ characterizes the
contribution (weight) of a bare particle state $a_p$ to the quasiparticle
state with the energy $E^{(i)}(p)$ . For this reason $\sum_iZ_p^{(i)}=1$.

First, if the energy of the antiferromagnetic exchange is the largest
energetic parameter $(I>>J,t)$ then the formation of the local polaron $b_r$
is energetically disadvantageous because this formation leads to the
disintegrating of antiferromagnetic correlations between spins $S_r$ and $%
S_{r+g}$. In Fig.\ref{f1} the quasiparticle spectra $E^{(i)}(p)$ are given
along the symmetry line $p=p_x=p_y$ for $I=5,J=t=1$. The local polaron bands
(they represent the motions of $b_r$ and $d_r$) lie essentially higher by
energy then two other bands which correspond to the hybridized spectra (\ref
{E}) with the gap $\Delta =J\sqrt{M}$. Two higher bands are characterized by
a small weight of spectral function $A(p,\omega )$ (see Fig. \ref{f1}(b)).
As to two lowest bands, the spectral function weight is close to 1 when the
band represents mainly the motion of a bare particle and is close to zero
for the Q-polaron band. In the region of strong hybridization, $p$ close to $%
\pi /2$, the spectral weight is of the same order for both bands. It may be
seen, that these two lowest bands have the symmetry close to the symmetry
given by the magnetic Brillouin zone, but the corresponding quasiparticles
have the full Brillouin zone symmetry .

Now let us discuss the second case, represented in Fig.\ref{f2} by the
spectrum for $J=I=0.1t$, when the spin interactions are small relative to
the kinetic energy of the electron, $t>>J,I$. In this case all
quasiparticles are approximately independent: the free motion of the bare
electron $B_{1p}=a_p$ is described by the band $E_1\cong \epsilon _p$ and
the antiferromagnetic polaron $B_{3p}$ moves in the band $E_3\cong \epsilon
_{p+Q}=-\epsilon _p$. As to the local polaron $B_{2p}$ it has it's own
dispersion $E_2\cong a(2,2)$, (\ref{M}). The absolute value of the
multiplier $(-\mid C_1\mid +M)/f_2$ of the term with $\epsilon _p$ in the
expression for $a(2,2)$ characterizes the essential band narrowing, i.e. the
increase of the local polaron mass relative to the mass of the bare
electron. Finally, the $B_{4p}$ polaron band $E_4\cong a(4,4)$ also
demonstrates a strong, but different renormalization of the mass. The
difference in the renormalization is one of the reasons why the resulting
hybridized spectra are asymmetrical relative to the magnetic Brillouin zone
boundary. The crossing of the nonhybridized spectra takes place close to
this boundary where $\epsilon _p=0.$ Out of this region one of the bands
explicitly represents the motion of the bare particle and has the spectral
weight $Z_p$ close to 1.

Fig.3 describes the case $J>>t,I$. The lowest bands (with the center at $%
E^{(s)}\cong -\frac 32J$) correspond to the motion of the local polaron in a
singlet state $b^{(s)}=b-\frac 12a$ and to the motion of such a polaron
coupled to the Q-wave. The upper bands (centered at $E^{(t)}\cong \frac 12J$%
) describe the motion of the local polaron in a triplet state $b^{(t)}=b+%
\frac 32a$ and also its coupling to the Q-wave. The ''singlet'' and
''triplet'' terminology suppose, that acting on the spin subsystem state,
the operators $b_r^{(s)+},b_r^{(t)+}$create accordingly a singlet or triplet
spin-electron pair in the site $r$.

The qualitative and important difference of this case from the previous ones
is the following. For two first cases the bottom of the lowest band is
determined mainly either by the motion of the bare particle or by the motion
of Q-polaron $c_p$ , the residues being close to 1 and zero correspondingly.
In the present case the bottom of the spectrum of the elementary excitations
turns out to be substantially determined by the $b_p$ and $d_p$-states. At
the same time, as it may be seen from Fig.\ref{f3}(b) (the solid line), the
possible filling of each $p$ state by a particle with a fixed spin is not
too small for the lowest band and is close to 0.1. Let us mind that such a
strong deviation of filling from 1 must lead to relatively great Fermi
surface even at small filling.

The present case demonstrates the importance of taking into account the
Q-polaron $d_p$. In order to illustrate this circumstance, in Fig. \ref{f4}
we show the spectrum of the elementary excitations calculated for the same
values of energetic parameters as in Fig.3, but using the basis of the first
three operators $a_p,b_p,c_p$. As it may be seen by comparison of Figs.3 and
4, this leads to the shift of the lowest band bottom from the point ${\bf p}%
=(0,0)$ to the point ${\bf p}=(\pi ,\pi )$. At the same time the value of
the residue $Z_p$ for the band bottom is close to 0.1 as in Fig.3.

Finally, if three energetic parameters are of the same order of magnitude
then the elementary excitation in any of four bands is a coherent
superposition of the states $a,b,c,d$ and these states have relatively the
same weight. It is notable that the presence of the long range order doesn't
lead to the periodicity of the spectrum relative to the magnetic Brillouin
zone.

Let us mind that the described above properties of the spin polaron are
mainly preserved if we suppose that the spin subsystem has no long range
order but the spin correlation length is large. This must lead to the
replacement of (\ref{ctw}) with the following relation which is true for $2D$
antiferromagnet at $T\not =0$:

$$
C_R^{\alpha \beta }=<S_r^\alpha S_{r+R}^\beta >=M^{\alpha \beta }e^{iQR}e^{-%
\frac RL},\quad L>>1. 
$$

In this case the delocalized polaron is described by the relation

$$
c_r=L^{-3}\sum_Re^{iQR}e^{-\frac RL}\tilde S_{r+R}a_r. 
$$

In conclusion we want to make several notes. First, above we investigated
the elementary excitations of the system - one electron plus the AFM spin
subsystem. It may be seen that all the results for the system with one hole
may be obtained from the discussed above case by substituting $(-t_g)$
instead of $t_g$ in the kinetic Hamiltonian.

Secondly, the ignoring of the Q-polarons, i.e., the states $c$ and $d$,
leads to the disappearance of bands, and can change essentially the
remaining bands. We think that it is important to take into account
Q-polarons in such models as $t-J$ and three-band Hubbard model, where the
particle motion is strongly coupled to the spin subsystem.

Finally, let us mention once more that our calculations are performed in the
limit of low filling. Nevertheless the calculated upper bands have physical
meaning if we shall study the electron transitions. In general case it may
be seen that the one-particle spectral density strongly depends on filling.

ACKNOWLEDGMENTS

We are grateful to L.B.Litinski for valuable discussions and comments. This
work was supported, in part, by the INTAS-PFBR (project No. 95-0591), by
RSFR (Grant No. 95-02-04239-a), by Russian National program on
Superconductivity (Grant No. 93080),ISI Foundation and EU NTAS Network
1010-CT930055.


\begin{figure}

\caption{(a)The quasipartical spectra $E^{(i)}(p)$ for $I=5\gg J=t=1$ along the
 symmetry line 
$p=p_x=p_y$; (b)  the values of the electron
 Green's function  $<<a_{p\sigma }\mid a^{+}_{p\sigma }>>$  residues 
$Z^{(i)}_p$ which correspond to the poles $E^{(i)}(p)$. Different line types 
represent four bands and
 corresponding residues.}

\label{f1}

\end{figure}

\begin{figure}
\caption{(a)The quasipartical spectra $E^{(i)}(p)$ for $t=1\gg I=J=0.1$ along the
 symmetry line 
$p=p_x=p_y$; (b)  the values of the electron
 Green's function  $<<a_{p\sigma }\mid a^{+}_{p\sigma }>>$  residues 
$Z^{(i)}_p$ which correspond to the poles $E^{(i)}(p)$, $Z^{(i)}_p$ are 
shown for the interval close to $p=\pi /2$. Different line types 
represent four bands and
 corresponding residues.}

\label{f2}

\end{figure}

\begin{figure}
\caption{(a)The quasipartical spectra $E^{(i)}(p)$ for $J=5\gg t,I,t=1,I=0.1$ along the
 symmetry line 
$p=p_x=p_y$; (b)  the values of the electron
 Green's function  $<<a_{p\sigma }\mid a^{+}_{p\sigma }>>$  residues 
$Z^{(i)}_p$ which correspond to the poles $E^{(i)}(p)$. Different line types 
represent four bands and
 corresponding residues.}

\label{f3}

\end{figure}

\begin{figure}
\caption{The same as in Fig.3, but for truncated basis of three operators 
$a_p, b_p, c_p$}

\label{f4}

 \end{figure}

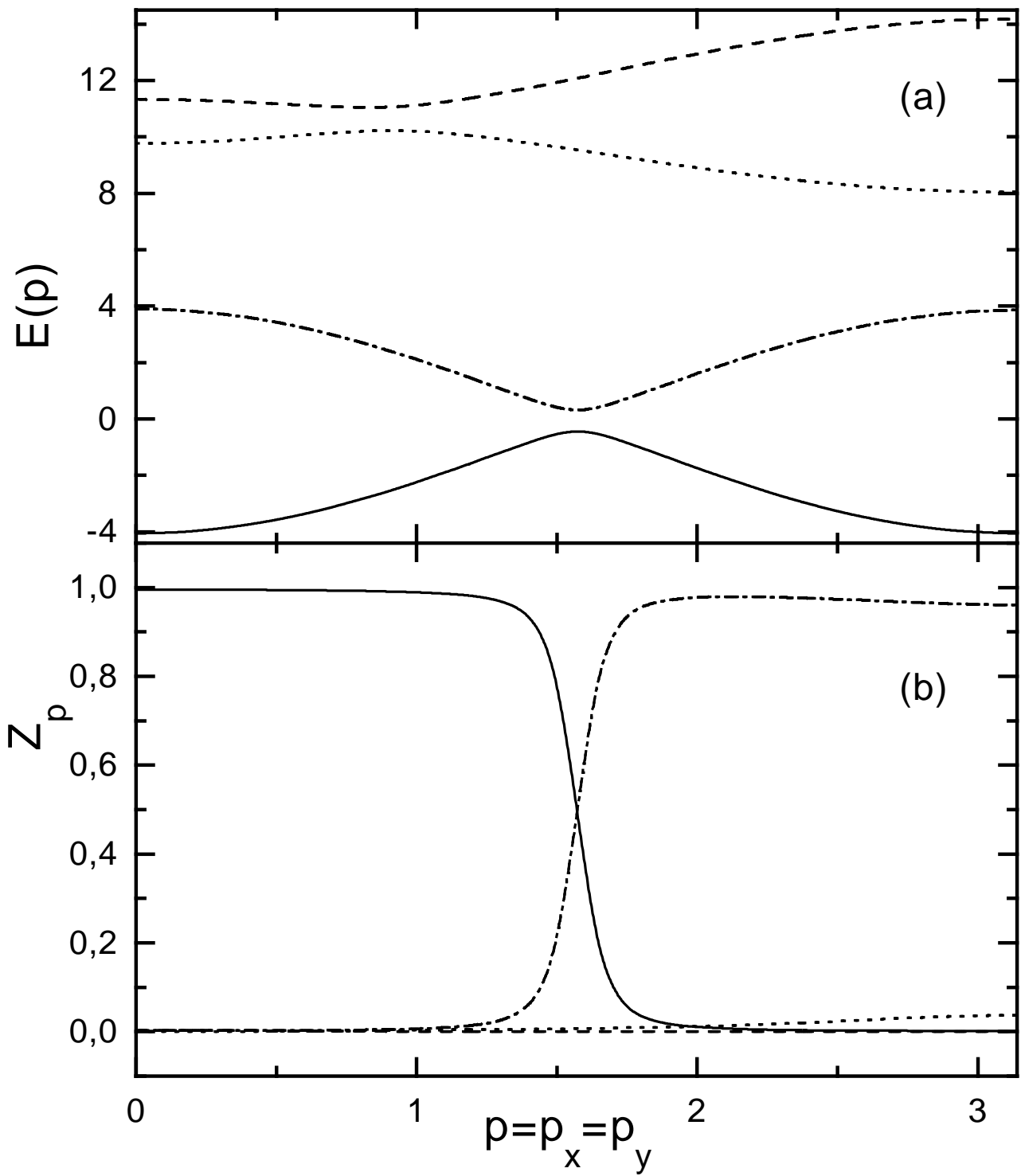
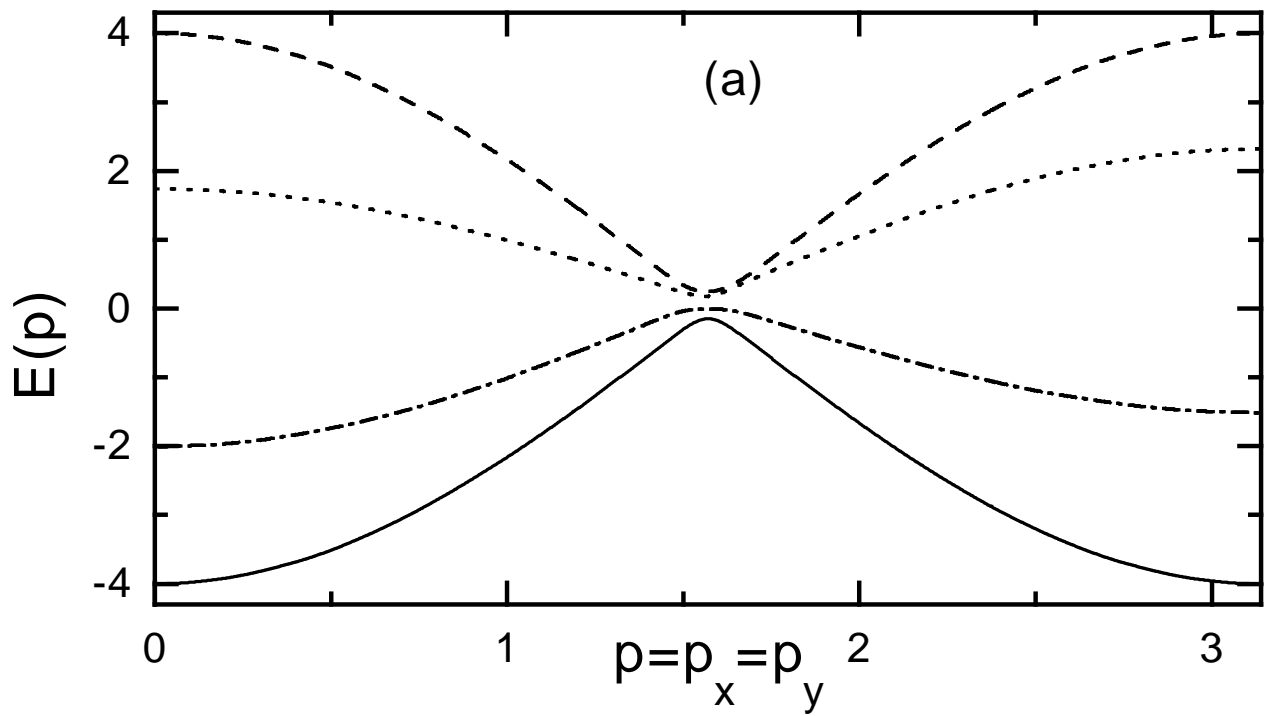
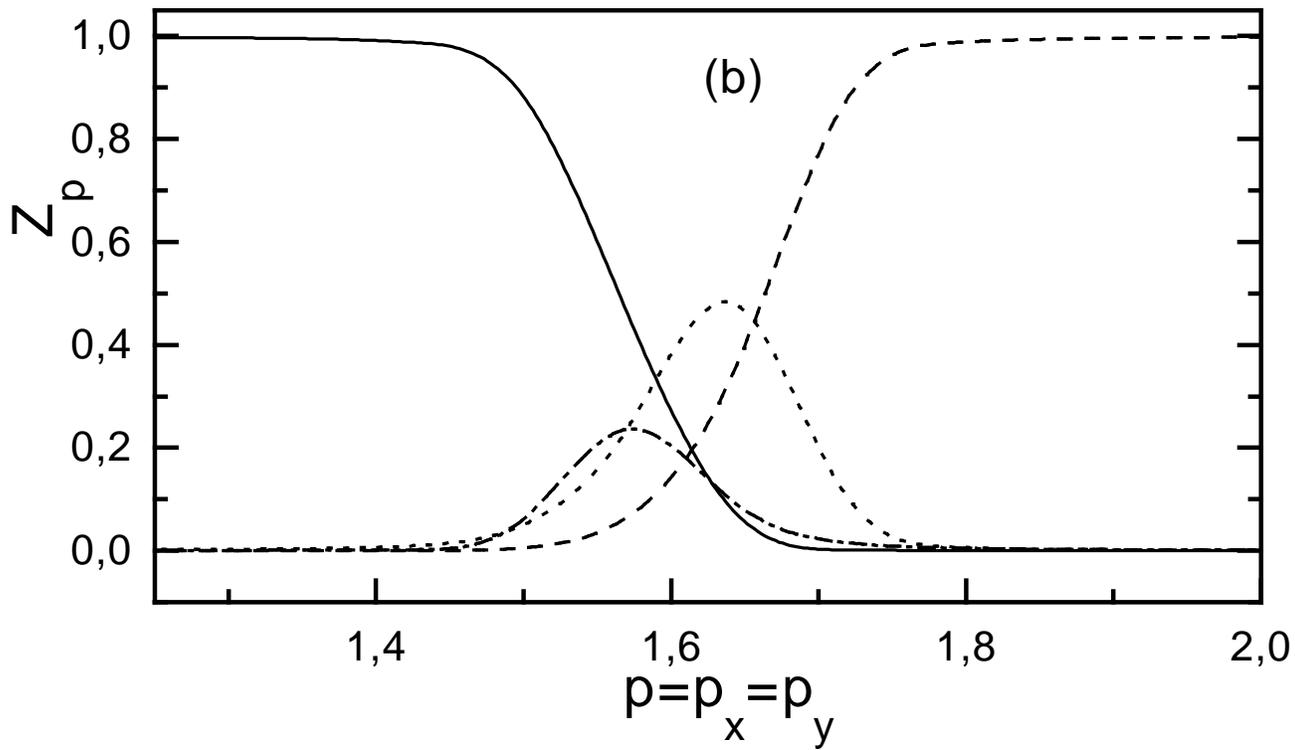
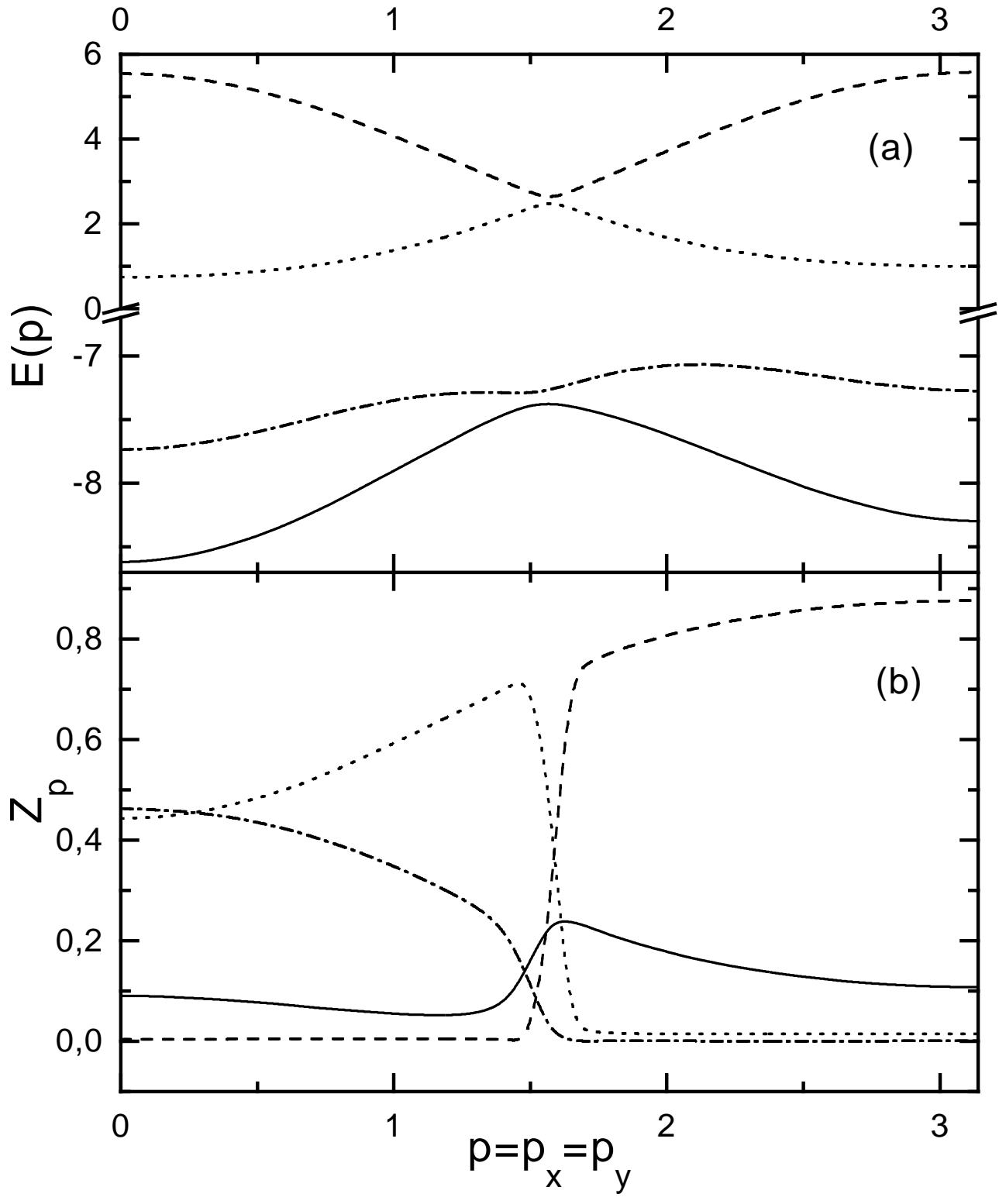
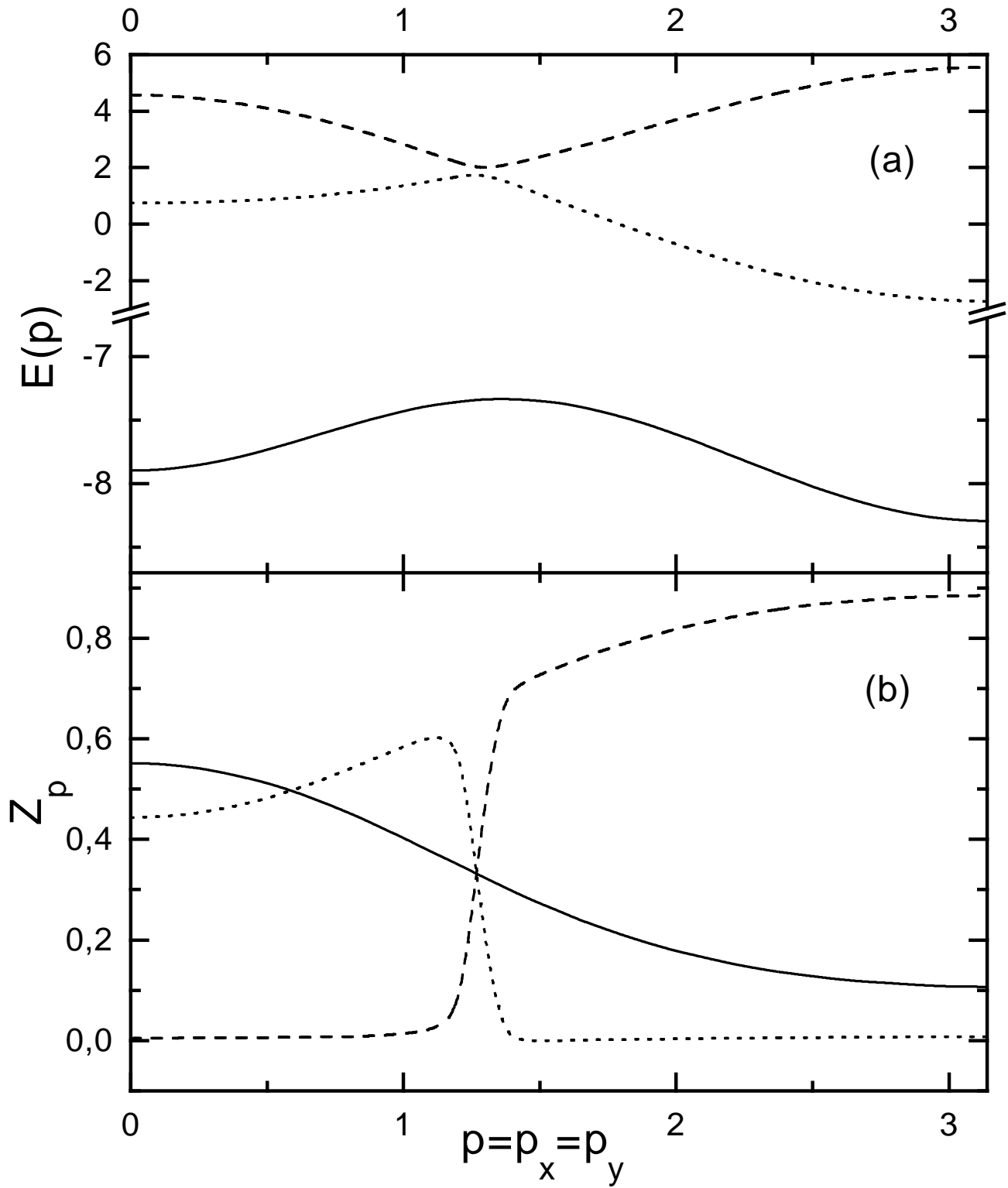
\end{document}